\documentclass[aps,pre,superscriptaddress,twocolumn,showpacs]{revtex4-1}

\usepackage{amsmath}
\usepackage{amsfonts}
\usepackage{graphicx}
\usepackage{epstopdf}
\usepackage{xcolor}

\begin{document}


\title{Clues on chemical mechanisms from renormalizability: The example of a noisy cubic autocatalytic model}


\author{Jean-S\'{e}bastien Gagnon}
\email[]{gagnon01@fas.harvard.edu}
\affiliation{Department of Earth and Planetary Sciences, Harvard University, Cambridge, Massachusetts, USA}

\author{Juan P\'{e}rez-Mercader}
\email[]{jperezmercader@fas.harvard.edu}
\affiliation{Department of Earth and Planetary Sciences, Harvard University, Cambridge, Massachusetts, USA}
\affiliation{Santa Fe Institute, Santa Fe, New Mexico, USA}


\date{\today}

\begin{abstract}
We study the effect of noise on the renormalizability of a specific reaction-diffusion system of equations describing a cubic autocatalytic chemical reaction.  The noise we are using is gaussian with power-law correlations in space, characterized by an amplitude $A$ and a noise exponent $y$.  We show that changing the noise exponent is equivalent to the substitution $d_{s} \rightarrow d_{\rm eff} = d_{s} - y$  and thus modifies the divergence structure of loop integrals ($d_{s}$ is the dimension of space).  The model is renormalizable at one-loop for $d_{\rm eff} < 6$ and nonrenormalizable for $d_{\rm eff} \geq 6$.  The effects of noise-generated higher order interactions are discussed.  In particular, we show how noise induces new interaction terms that can be interpreted as a manifestation of some (internal) ``chemical mechanism''.  We also show how ideas of effective field theory can be applied to construct a more fundamental chemical model for this system.
\end{abstract}

\pacs{05-10-Cc, 82-20-w, 02-50-r}

\maketitle

\section{Introduction}

The nature of chemical reaction mechanisms is a scale dependent problem of great practical and theoretical importance \cite{Steinfeld_etal_1998}, often studied using spectroscopic or other analytical (in the chemical sense) techniques.  A key question in chemistry and biology is to understand the mechanisms and chemical reactions behind the macroscopic behavior of complex systems such as cells.  Answering the above question is of course a daunting task.  However, given the ubiquity and relevance of these systems, it is important to try to gain insight from simpler, more tractable models.

In this paper we approach the question of inferring the small scale (ultraviolet, UV) structure of a system known at large scales (infrared, IR) in the context of stochastic reaction-diffusion equations.  In other words, we want to fine-grain those equations in order to study the underlying mechanisms from the large scale dynamics.  Although they are relatively simple, reaction-diffusion equations lead to complex patterns~\cite{Cross_Hohenberg_1993} and are thought to be an essential part of morphogenesis~\cite{Turing_1952,Meinhardt_1982,Tompkins_etal_2014}.  In addition to serve as models for biological pattern formation, they are also used in various contexts such as the spreading of epidemics~\cite{Murray_etal_1986}, ecological invasions~\cite{Holmes_etal_1994}, tumour growth~\cite{Lowengrub_etal_2010} and oscillating chemical reactions~\cite{Cross_Hohenberg_1993,Epstein_Pojman_1998}. 

For definiteness, we focus on a particular reaction-diffusion model~\cite{Pearson_1993,Lesmes_etal_2003} based on a cubic autocatalytic two-species system~\cite{Higgins_1964,Selkov_1968,Gray_Scott_1983,Gray_Scott_1984, Gray_Scott_1985,Prigogine_Nicolis_1977}.  This simple model has a very interesting phenomenology.  Numerical simulations of the deterministic \cite{Pearson_1993} and stochastic \cite{Lesmes_etal_2003} versions of the model show the formation of domains (``cells'').  These domains share some characteristics with living systems like birth, growth, movement, replication and death. Understanding the inner dynamics of such a simple system might provide insight into real, more complex organisms.

The analysis of stochastic reaction-diffusion equations can be couched in the language of field theory (e.g. \cite{Barabasi_Stanley_1995}).  In the following, we outline the first steps in a new approach to study the above cubic autocatalytic reaction-diffusion (CARD) model, that could in principle be applied to other reaction-diffusion systems.  A hallmark of our method is the use of a fine-graining strategy more akin to the philosophy in particle physics, in contrast to the coarse-graining methods typically used in condensed matter physics.  In other words, we apply the renormalization group to the CARD model, but run it from the IR to the UV.  The latter strategy is natural in the context of finding more fundamental microscopic models from the knowledge of a macroscopic one. 

Another important ingredient in our approach is the use of noise to represent the combined contributions of both internal degrees of freedom and environmental effects.  We argue that we may use externally tunable noise to probe reaction-diffusion equations at small scales, in the same way that varying beam energy is used to study the inner structure of particles.

The above approach has been applied to the study of small-scale structures in the CARD model \cite{Gagnon_etal_2015}.  In the present contribution, we develop the more general aspects of the approach.  The rest of the paper is organized as follows.  In Sect.~\ref{sec:Philosophy_goals}, we present the general idea behind our fine-graining method and its potential relevance to reaction-diffusion equations and chemistry.  In particular, we show how self-consistent approaches (i.e. renormalizability, effective interactions, decoupling, see for example \cite{AlvarezGaume_2012,Schwartz_2014,Burgess_2007,Appelquist_Carazzone_1974,Weisberger_1981,Galvan_etal_1987}) could allow the analysis of the system's inner structures.  The specific details of the CARD model are introduced in Sect.~\ref{sec:CARD_model}.  Section~\ref{sec:Renormalizability} presents the proof of renormalizability of the CARD model at one-loop, and paves the way for the introduction of higher order interactions (Sect.~\ref{sec:Higher_order_interactions}) and effective field theory (Sect.~\ref{sec:EFT_chemistry}).  We discuss our results in Sect.~\ref{sec:Discussion}.

\section{Philosophy and goals}
\label{sec:Philosophy_goals}

The goal of this section is to explain our method for studying underlying mechanisms in a broader setting.  As an example, let us take the following macroscopic chemical equation:
\begin{eqnarray}
\label{eq:Example_macroscopic}
2\mbox{H}_{2} + \mbox{O}_{2} & \stackrel{k}{\rightarrow} & 2\mbox{H}_{2}\mbox{O}
\end{eqnarray}
where $k$ is the (forward) reaction rate.  This chemical equation is valid at large temporal and spatial scales (corresponding to low momentum or resolution).  At shorter temporal and spatial scales, the above description might break down and must be replaced by another one.  One example of microscopic description is:
\begin{eqnarray}
\label{eq:Example_microscopic_1}
\mbox{O}_{2} & \stackrel{k_{1}}{\rightarrow} & 2\mbox{O} \nonumber \\
2\mbox{H}_{2} + 2\mbox{O} & \stackrel{k_{2}}{\rightarrow} & 2\mbox{H}_{2}\mbox{O}
\end{eqnarray}
Another possible microscopic description is:
\begin{eqnarray}
\label{eq:Example_microscopic_2}
\mbox{H}_{2} + \mbox{O}_{2} & \stackrel{k_{1}}{\rightarrow} & 2\mbox{OH} \nonumber \\
\mbox{H}_{2} & \stackrel{k_{2}}{\rightarrow} & 2\mbox{H} \nonumber \\
2\mbox{H} + 2\mbox{OH} & \stackrel{k_{3}}{\rightarrow} & 2\mbox{H}_{2}\mbox{O}
\end{eqnarray}
Other descriptions (or underlying mechanisms) are possible.  We do not know a priori the underlying mechanism of this reaction, some experimental input is needed.  The correct (and very complicated) mechanism can be found in Ref.~\cite{Steinfeld_etal_1998}.  To find the right microscopic description, one needs to directly measure the microscopic reactions using very high resolution probes or use indirect techniques~\cite{Steinfeld_etal_1998,Ross_2008}.  This is often an experimental challenge.

The same could be said of particle physics.  To find the correct microscopic particle physics model, one needs to directly detect and characterize the new particles using high-energy accelerators.  This is difficult, and requires very high resolution probes.  In the case where the scale of new physics is very high, it is not even possible to directly probe these new particles as the energies required are not directly accessible to the accelerators.  Fortunately, the framework of effective field theory enables one to find clues about the microscopic model by studying the parameters of the macroscopic model (e.g. \cite{Burgess_2007}).  

The idea of applying noise (or fluctuations) to a system to get information about its components is not new (e.g. \cite{Ross_2008}).  The effect of fluctuations on the behavior of a system depends (among other things) on the size of the system and the strength and type of those fluctuations.  Renormalization provides us with a set of tools that allows to compute the change in a model's parameters due to fluctuations as a function of scale.  Thus in analogy to particle physics, we argue in the following how a combination of externally tunable noise and renormalization techniques in the context of effective field theory can be used to study underlying chemical mechanisms.

There are two ways to apply the renormalization group: coarse-graining (i.e. from the UV to the IR, typically used in condensed matter physics) or fine-graining (i.e. from the IR to the UV, typically used in particle physics).  In condensed matter, a small momentum shell is integrated at a large momentum cutoff $\Lambda$, resulting in a differential equation giving the running of the model parameters from small to large scales (coarse-graining).  The cutoff $\Lambda$ (e.g. inverse lattice spacing) is considered to be physical, along with all model parameters (see for example \cite{Medina_etal_1989,Frey_tauber_1994} for an application of this method to the Kardar-Parisi-Zhang equation and~\cite{Hochberg_etal_2003} for an application to the CARD model).  In particle physics, the cutoff $\Lambda$ is typically very large and not known a priori, so parameters are renormalized in such a way as to make them independent of $\Lambda$.  Since none of the model parameters are physical from the outset, they are given physical meaning by experiments at a certain scale (typically large), from which the running towards shorter distance scales proceeds.  Both approaches are in principle equivalent, but differ in their practical implementation  See Refs.~\cite{Binney_etal_2001,Peskin_Schroeder_1995} for a comparison of the two approaches.

The renormalization group allows to compute the change in parameters from a low momentum scale $\Lambda_{0}$ to a high momentum scale $\Lambda$, provided there are no additional dynamical degrees of freedom appearing when going from $\Lambda_{0}$ to $\Lambda$ (or vice-versa).  This is clearly not the case when going from Eq.~(\ref{eq:Example_macroscopic}) (3 dynamical degrees of freedom H$_{2}$, O$_{2}$ and H$_{2}$O) to Eq.~(\ref{eq:Example_microscopic_1}) (4 dynamical degreees of freedom H$_{2}$, O$_{2}$, O and H$_{2}$O) or Eq.~(\ref{eq:Example_microscopic_2}) (5 dynamical degrees of freedom H$_{2}$, H, O$_{2}$, OH and H$_{2}$O).  Assume there is an intermediate scale $\Lambda_{i}$ at which new dynamical degrees of freedom are introduced.  Then the renormalization group can be run from $\Lambda_{0}$ to $\Lambda_{i}$  using the initial degrees of freedom, and from $\Lambda_{i}$ to $\Lambda$ using the additional degrees of freedom.  The two renormalization group flows must match at $\Lambda_{i}$.

The main goal of this paper is to find clues about mechanisms acting at momenta larger than $\Lambda_{i}$.  Information about high momenta processes manifests itself as corrections in the parameters of the model due to irrelevant terms.  This can be seen as follows.  We can write down the rate equation for H$_{2}$ corresponding to Eq.~(\ref{eq:Example_macroscopic}):
\begin{eqnarray}
\label{eq:Rate_equation_macroscopic}
\frac{d[\mbox{H}_{2}]}{dt} & = & -k[\mbox{H}_{2}]^{2}[\mbox{O}_{2}] - k_{i}[\mbox{H}_{2}]^{p}[\mbox{O}_{2}]^{q}
\end{eqnarray}
with $p + q > 3$ and where $[...]$ means concentration.  The first term on the RHS corresponds to the expected chemical rate law and the second term is higher order in the reactant concentrations.  Following Polchinski~\cite{Polchinski_1984}, we write the renormalization group flows of the two parameters as:
\begin{eqnarray}
\label{eq:RG_flow_example_1}
\mu\frac{dg}{d\mu} & = & \beta(g,\mu^{2}g_{i}) \\
\label{eq:RG_flow_example_2}
\mu\frac{dg_{i}}{d\mu} & = & \mu^{-2}\beta_{i}(g,g_{i})
\end{eqnarray}
where $\mu$ is a momentum scale, $g \propto O(k)$ is dimensionless and $g_{i} \propto O(k_{i})$ has negative momentum dimensions (assumed to be $\mu^{-2}$ for illustrative purposes).  Defining the dimensionless variables $\lambda = g$ and $\lambda_{i} = \mu^{2}g_{i}$, Eqs.~(\ref{eq:RG_flow_example_1})-(\ref{eq:RG_flow_example_2}) become:
\begin{eqnarray}
\label{eq:RG_flow_example_dimensionless_1}
\mu\frac{d\lambda}{d\mu} & = & \beta(\lambda,\lambda_{i}) \\
\label{eq:RG_flow_example_dimensionless_2}
\mu\frac{d\lambda_{i}}{d\mu} - 2\lambda_{i} & = & \beta_{i}(\lambda,\lambda_{i})
\end{eqnarray}
Taking a particular solution to both equations ($\tilde{\lambda}$, $\tilde{\lambda}_{i}$) and considering small deviations from these solutions ($\Delta = \lambda - \tilde{\lambda}$, $\Delta_{i} = \lambda_{i} - \tilde{\lambda}_{i}$), we can write to first order in the deviations:
\begin{eqnarray}
\label{eq:RG_flow_example_deviations_1}
\mu\frac{d\Delta}{d\mu} & = & \frac{\partial\beta(\tilde{\lambda},\tilde{\lambda}_{i})}{\partial\lambda}\Delta + \frac{\partial\beta(\tilde{\lambda},\tilde{\lambda}_{i})}{\partial\lambda_{i}}\Delta_{i} \\
\label{eq:RG_flow_example_deviations_2}
\mu\frac{d\Delta_{i}}{d\mu} - 2\Delta_{i} & = & \frac{\partial\beta_{i}(\tilde{\lambda},\tilde{\lambda}_{i})}{\partial\lambda}\Delta + \frac{\partial\beta_{i}(\tilde{\lambda},\tilde{\lambda}_{i})}{\partial\lambda_{i}}\Delta_{i}
\end{eqnarray}
The $-2\Delta_{i}$ term in Eq.~(\ref{eq:RG_flow_example_deviations_2}) implies that the deviation $\Delta_{i}$ is strongly damped when following the flow from large to small momenta.  In the present case, the suppression factor is $O(\mu^{2}/\Lambda_{i}^{2})$.  There is no such damping for $\Delta$.  We thus conclude that the first term on the RHS of Eq.~(\ref{eq:Rate_equation_macroscopic}) is relevant, while the second term is irrelevant.  Note that power-law damped irrelevant terms correspond to nonrenormalizable terms in effective field theory language \cite{Burgess_2007,Binney_etal_2001,Peskin_Schroeder_1995}.

The damping depends in general on the engineering dimension of the parameter considered.  We show in Sect.~\ref{sec:Renormalizability} how the exponent of power-law noise affects this engineering dimension and thus the (non) renormalizability of terms in the CARD model.  The damping also depends on $\Lambda_{i}$, the momentum scale at which new chemical reactions (underlying mechanisms) are taking place.  A large (small) value of $\Lambda_{i}$ implies a strong (weak) suppression of the irrelevant terms at low momenta.  The size of irrelevant terms is thus a measure of how sensitive the system is to underlying mechanisms.

The effect of irrelevant (nonrenormalizable) terms on experimentally accessible parameters can be seen from Eqs.~(\ref{eq:RG_flow_example_deviations_1})-(\ref{eq:RG_flow_example_deviations_2}).  Before $\Delta_{i}$ is damped out at low momenta, it affects the behavior of $\Delta$ through the second term on the RHS of Eq.~(\ref{eq:RG_flow_example_deviations_1}).  Thus irrelevant terms add corrections (suppressed by powers of $\Lambda_{i}$) to the relevant parameters of the model.

Experimentally, it is in principle possible to measure the effects of underlying mechanisms from variations in the parameters of the model.  Indeed, externally tunable noise is an experimental handle on the system that helps ``break the system apart'' and bring nonrenormalizable terms to the forefront.  Effective field theory tools (developed in the context of reaction-diffusion equations in the rest of this paper) give a framework to compute those corrections.

\section{The stochastic CARD model}
\label{sec:CARD_model}

In order to develop some intuition for our fine-graining approach applied to reaction-diffusion equations, we consider a stochastic version of the CARD model~\cite{Lesmes_etal_2003} (note that the tools developed in the following sections are in principle applicable to other reaction-diffusion systems).  It is based on the following chemical reactions~\cite{Gray_Scott_1983,Gray_Scott_1984,Gray_Scott_1985}:
\begin{eqnarray}
\label{eq:Gray_Scott_reactions}
\mbox{U + 2V} & \stackrel{\lambda}{\rightarrow} & 3\mbox{V},\;\;\;\;\; \mbox{V} \stackrel{r_{v}}{\rightarrow}\;\; ,\;\;\;\;\; \mbox{U} \stackrel{r_{u}}{\rightarrow}\;\; ,\;\;\;\;\;  \stackrel{f}{\rightarrow} \mbox{U}.
\end{eqnarray}
A possible interpretation is to view $\mbox{U}$ as food and $\mbox{V}$ as an organism building its own body parts from food (a crude form of metabolism).  The evolution equations corresponding to reactions~(\ref{eq:Gray_Scott_reactions}) in the general case where diffusion and noise are present are:
\begin{eqnarray}
\label{eq:Gray_Scott_equations_1}
\partial_{t} V & = & D_{v}\nabla^{2}V - r_{v}V + \lambda UV^{2} + \eta_{v}(x), \\
\label{eq:Gray_Scott_equations_2}
\partial_{t} U & = & D_{u}\nabla^{2}U  - r_{u}U - \lambda UV^{2}  + f + \eta_{u}(x),
\end{eqnarray}
where we use the shortcut notation $x = ({\bf x},t)$, $B=B(\vec{x},t)$ is the spacetime dependent concentration for species B (with B = U,V), $D_{b}$ is the diffusion constant for species B, $r_{b}$ is the decay rate into inert products for species B, $\eta_{b}(x)$ is the spacetime dependent stochastic noise term for species B, $\lambda$ is the rate constant for the autocatalytic reaction between U and V and $f$ is the constant feed rate of U into the system.  All model parameters are positive.  When $U \gg  2f/r_{u}$, feeding more U does not produce significant effects on the amount already present and the $f$ term in Eq.~(\ref{eq:Gray_Scott_equations_2}) can be neglected.  For simplicity, we assume that we are in such a regime for the rest of this paper.  Note that the feeding term $f$ can always be reintroduced into the analysis in the form of additional Feynman diagrams.

\begin{figure}
\includegraphics[width=0.48\textwidth]{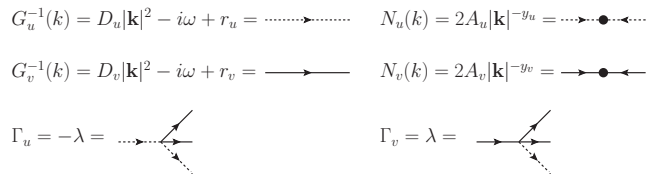}
\caption{Summary of Feynman rules.  See the discussions below Eqs.~(\ref{eq:Gray_Scott_equations_2}) and (\ref{eq:Noise_property_4}) for symbol definitions. \label{fig:Feynman_rules}}
\end{figure}

In Fourier space, Eqs.~(\ref{eq:Gray_Scott_equations_1})-(\ref{eq:Gray_Scott_equations_2}) become two coupled equations that can be solved perturbatively provided a dimensionless combination of parameters involving $\lambda^{L + \frac{E}{2}-1}A_{b}^{L}$ is small.  Here $A_{b}$ is a noise amplitude (see below) and $L$, $E$ are the number of loops and external legs of Feynman diagrams obtained in the perturbative expansion of response functions.  Feynman rules for stochastic partial differential equations are well-known and discussed in Ref.~\cite{Barabasi_Stanley_1995}.  The specific rules we use are derived in Ref.~\cite{Hochberg_etal_2003}: they include the components shown in Fig.~\ref{fig:Feynman_rules}, supplemented with momentum conservation at each vertex and integration over undetermined momenta.

The noise term in Eqs.~(\ref{eq:Gray_Scott_equations_1})-(\ref{eq:Gray_Scott_equations_2}) needs justification.  First, noise plays an important role in physics~\cite{MacDonald_2006,Dutta_Horn_1981}, chemistry~\cite{vanKampen_2007,Gillespie_2007} and biology~\cite{Tsimring_2014,Bajic_Wee_2005}, since it can be used to model fluctuations due to the intrinsic (quantum mechanical) randomness of a system or external effects due to the environment (such as temperature, luminosity, stirring, etc).  In the following, we focus on additive external noise, since internal noise is negligible for a macroscopic system ($U \gg  2f/r_{u}$) at relevant experimental temperatures \footnote{The treatment of internal noise requires more refined tools, as discussed in the review~\cite{Tauber_etal_2005}.  See also Ref.~\cite{Cooper_etal_2013} for a study of internal noise in the CARD model).}.

From the above considerations, there are two ways to interpret the additive external noise terms in Eqs.~(\ref{eq:Gray_Scott_equations_1})-(\ref{eq:Gray_Scott_equations_2}).  The first way is to consider it as an environmental effect that has a direct influence on the evolution of the system.  For example, it is shown in Ref.~\cite{Lesmes_etal_2003} that noise amplitude controls the type of patterns (no pattern, stripes, replicating and non-replicating cell-like domains) appearing in the stochastic CARD model.  The second way is to use noise as an external tool to manipulate the system and probe its microscopic dynamics.  In this interpretation, the noise parameters are controlled experimentally and free to take any value.  The latter is the focus of the present paper.

In the following, we use a gaussian noise with power-law correlations (c.f.  \cite{Hochberg_etal_2003}):
\begin{eqnarray}
\label{eq:Noise_property_1}
\langle \eta_{u}(k) \rangle & = & \langle \eta_{v}(k) \rangle \;\;\;=\;\;\; 0, \\
\label{eq:Noise_property_2}
\langle \eta_{v}(k)\eta_{v}(p) \rangle & = & 2A_{v}|{\bf k}|^{-y_{v}}(2\pi)^{d_{s}+1}\delta^{(d_{s}+1)}(k+p), \\
\label{eq:Noise_property_3}
\langle \eta_{u}(k)\eta_{u}(p) \rangle & = & 2A_{u}|{\bf k}|^{-y_{u}}(2\pi)^{d_{s}+1}\delta^{(d_{s}+1)}(k+p), \\
\label{eq:Noise_property_4}
\langle \eta_{v}(k)\eta_{u}(p) \rangle & = & \langle \eta_{u}(k)\eta_{v}(p) \rangle \;\;\;=\;\;\; 0,
\end{eqnarray}
where we use the shortcut notation $k = ({\bf k},\omega)$ and we have expressed the correlations in Fourier space for later convenience.  $d_{s}$ is the dimension of space.   All higher order moments are zero.  The noise amplitudes $A_{b} > 0$ and exponents $y_{b}$ are free parameters of the model.  The motivation behind our choice of power spectrum for the noise is the following.  Power laws are found in many natural and man-made systems, and there exist various plausible mechanisms to produce them~\cite{Newman_2005}.  Self-organized criticality~\cite{Bak_etal_1987} is an example of mechanism that produces $1/f$ noise naturally.  The power-law spectrum~(\ref{eq:Noise_property_1})-(\ref{eq:Noise_property_4}) includes white noise ($y_{b} = 0$) as a special case and allows for stronger fluctuations and possibly more complex, scale-dependent patterns.  Another motivation behind the use of power laws is that they can be used as a basis to Taylor-expand more complex noise functions.

\section{Renormalizability of the stochastic CARD model}
\label{sec:Renormalizability}

From a phenomenological point of view, the question of renormalizability is important in field theory.  In general, perturbative solutions typically lead to UV or IR divergences (we focus on UV divergences in the following, since we are interested in going from large to small spatial/temporal scales).  Those divergences are hidden in loop diagrams, where integration over undetermined momenta is implicit.  One important factor determining the degree of divergence is the dimension of spacetime.  The usual program of renormalization is to absorb those divergences in the parameters of the model, leading to their dependence (or ``running'') with scale (e.g.~\cite{AlvarezGaume_2012}).  A renormalizable model (in the UV) has a finite number of UV divergences that are absorbed in a finite number of parameters.  Such a model is predictive in the sense that it is applicable all the way down to the smallest scales, provided no new physics is present at shorter scales.  In contrast, a nonrenormalizable model has an infinite number of divergences and is valid only down to a certain scale, where it must be replaced by a more fundamental one.  

Since we are interested in short distance dynamics, we need to explore the connection between noise and renormalizability.  To do that we use power counting arguments on loop corrections to the parameters of the model.  A generic one-loop integral with $n+1$ response functions has the form:
\begin{equation}
\label{eq:Generic_loop_integral}
I(n,d_{s},y) = \int\frac{d^{d_{s}}p}{(2\pi)^{d_{s}}}\; \frac{|{\bf p}|^{-y_{b}}}{(|{\bf p}|^{2} + \Delta^{2})^{n}} \sim \Lambda^{d_{s}-y_{b}-2n},
\end{equation}
where $\Delta$ is some dimensional quantity and we take all external momenta to be zero (this is sufficient for power counting purposes).  The last step gives the leading UV behavior of the integral with cutoff $\Lambda$.  Note that the noise introduces extra powers of momentum in Eq.~(\ref{eq:Generic_loop_integral}), and thus the divergence structure of the loop integral depends on the effective dimension $d_{\rm eff} \equiv d_{s} - y_{b}$.  Since $y_{b}$ is an external parameter, it implies that $d_{\rm eff}$ can be changed arbitrarily, giving rise to (possibly fractional) divergences.  Varying the noise exponent $y_{b}$ has important consequences, such as changing the relevance of interactions in the evolution equations.  We explore this situation in the following.  For later convenience we define $d_{\rm eff} \equiv 2m_{b} + \epsilon$, where $m_{b} \in \mathbf Z_{>0}$ and $0 \leq \epsilon < 2$.

\begin{figure}[ht]
\includegraphics[width=0.46\textwidth]{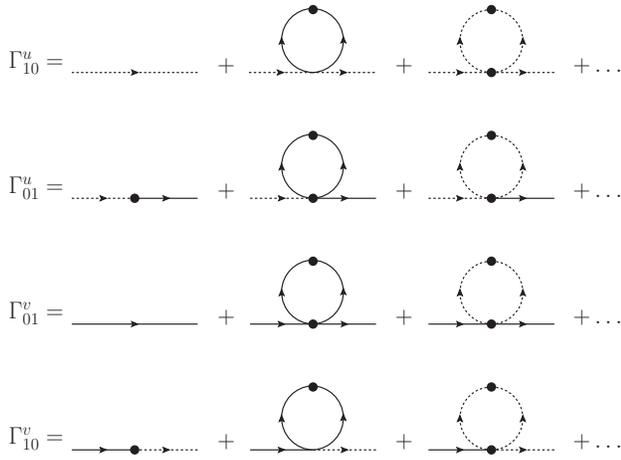}
\caption{Sample of one-loop 2-point functions.  A blob on a vertex represents a vertex that is not present at tree-level.   \label{fig:2_point_functions}}
\end{figure}

\begin{figure}[ht]
\includegraphics[width=0.46\textwidth]{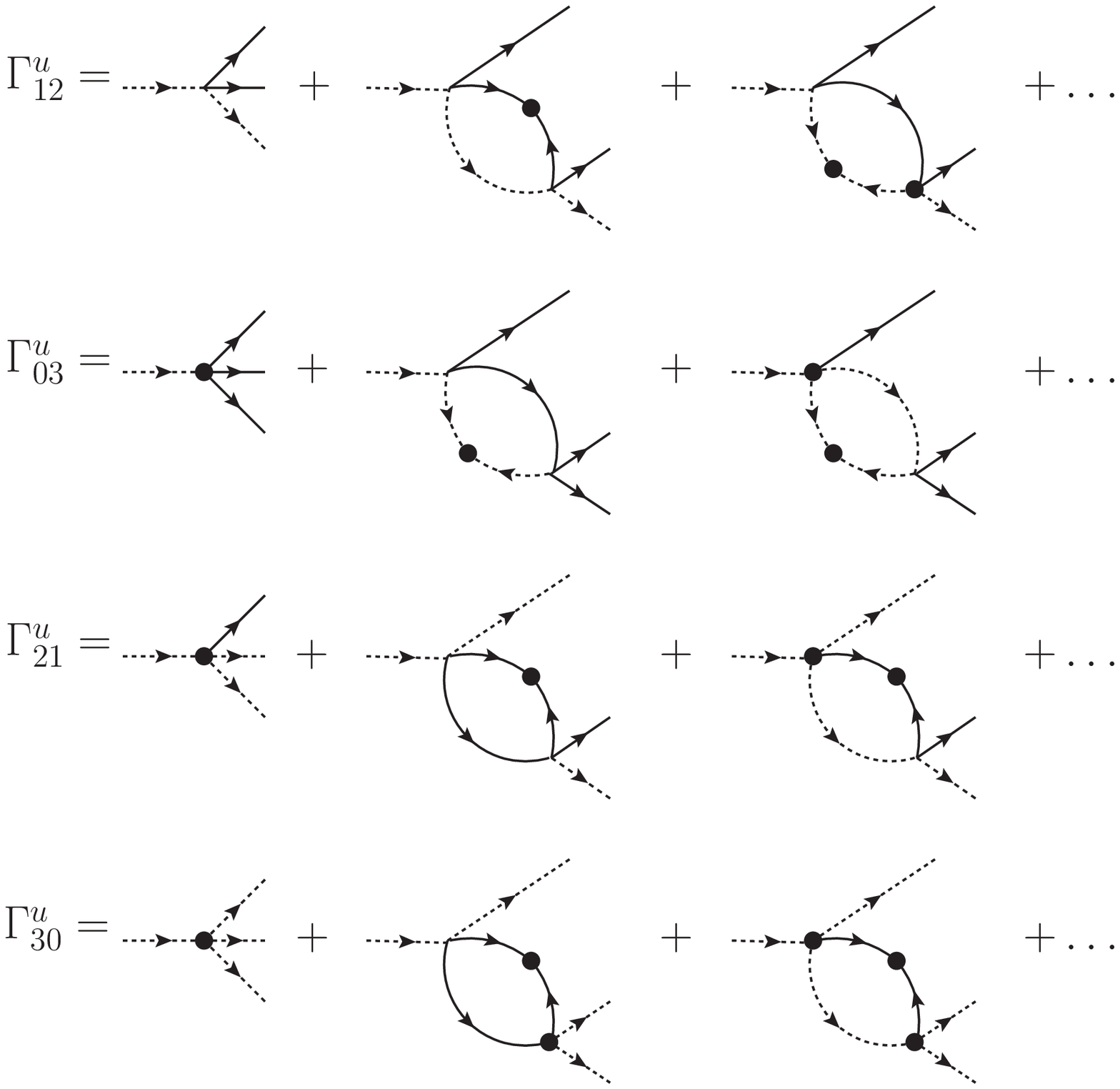}
\caption{Sample of one-loop 4-point functions.  A blob on a vertex represents a vertex that is not present at tree-level. \label{fig:4_point_functions}}
\end{figure}

\begin{figure}[ht]
\includegraphics[width=0.46\textwidth]{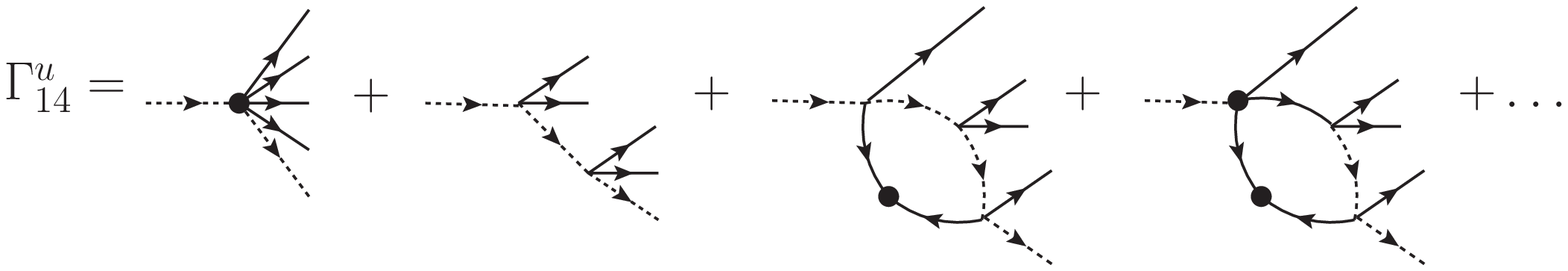}
\caption{Sample of one-loop 6-point functions.  A blob on a vertex represents a vertex that is not present at tree-level. \label{fig:6_point_functions}}
\end{figure}

A sample of 2, 4 and 6-point functions is shown in Figs.~\ref{fig:2_point_functions}-\ref{fig:6_point_functions}.  We use the notation $\Gamma_{pq}^{b}$ for $(1+p+q)$-point functions, where $b$ denotes the incoming chemical field and $p$, $q$ the number of outgoing $U$'s, $V$'s respectively.  The UV behaviors of typical 2, 4 and 6-point functions are:
\begin{eqnarray}
\Gamma_{10}^{u}(0) & \sim & r_{u} + \lambda A_{v}\sum_{j=1}^{m_{v}} C_{10}^{(j)}\Lambda^{2j+\epsilon -2} + O(\lambda_{30}),\\
\Gamma_{12}^{u}(0) & \sim & \lambda + \lambda^{2}A_{v} \sum_{j=1}^{m_{v}} C_{12}^{(j)}\Lambda^{2j+\epsilon -4} + O(\lambda\lambda_{21}),  \\
\Gamma_{14}^{u}(0) & \sim & \lambda_{14} + \lambda^{3}A_{u} \sum_{j=1}^{m_{u}} C_{14}^{(j)}\Lambda^{2j+\epsilon -6} + O(\lambda^{2}\lambda_{03}).
\end{eqnarray}
where $C_{pq}$ is a dimensionful factor constructed from the dimensionful parameters of the model (diffusion constants and decay rates) and we define new (non-tree level) couplings as $\lambda_{pq}U^{p}V^{q}$.  Note that there is no correction to diffusion constants at one-loop (in the $U \gg  2f/r_{u}$ approximation).  

The divergence structure for each $\Gamma_{pq}^{b}$ depends on the noise exponent.  We study three representative cases  with $m_{b} = 1,2,3$ (higher values of $m_{b}$ are done in a similar way and lower values  do not lead to UV divergences).  For $m_{b} = 1$, $\Gamma_{10}^{u}$ is $\Lambda^{\epsilon}$ divergent while $\Gamma_{12}^{u}$ and $\Gamma_{14}^{u}$ are both finite in the $\Lambda \rightarrow \infty$ limit.  The $\Lambda^{\epsilon}$ divergence can be absorbed in $r_{u}$ following the usual program of renormalization, resulting in a scale-dependent decay rate.  Since all 4-point functions are finite, it is not necessary to introduce new couplings ($\lambda_{03}$, $\lambda_{21}$, $\lambda_{30}$) to absorb one-loop divergences.  This implies that all diagrams containing $\lambda_{03}$, $\lambda_{21}$, $\lambda_{30}$ (represented by blob vertices in Fig.~\ref{fig:4_point_functions}) are zero for this value of the noise exponent.  The same reasoning applies to 6-point (and higher point) functions.  There is thus a finite number of divergences (contained in $\Gamma_{10}^{b}$ and $\Gamma_{01}^{b}$) that can be absorbed in a finite number of parameters at this loop order in perturbation theory.

For $m_{b} = 2$, $\Gamma_{10}^{u}$ is $\Lambda^{2+\epsilon}$ divergent and $\Gamma_{12}^{u}$ is $\Lambda^{\epsilon}$ divergent, while $\Gamma_{14}^{u}$ is finite.  Those two divergences can be absorbed in $r_{u}$ and $\lambda$.  In addition to $\Gamma_{12}^{u}$, the other 4-point functions shown in Fig.~\ref{fig:4_point_functions} are also $\Lambda^{\epsilon}$ divergent.  It is therefore necessary to introduce three new nonzero tree-level couplings ($\lambda_{03}$, $\lambda_{21}$, $\lambda_{30}$) to absorb those divergences.  They correspond to new interaction terms in Eqs.~(\ref{eq:Gray_Scott_equations_1})-(\ref{eq:Gray_Scott_equations_2}) (e.g. $\lambda_{21}U^{2}V$) and thus produce additional subleading contributions to the parameters (represented by Feynman diagrams with blob vertices in Figs.~\ref{fig:2_point_functions}-\ref{fig:4_point_functions}).  Since higher point functions do not diverge, the {\em finite} number of divergences can be absorbed in the parameters of the model.

\begin{figure}
\includegraphics[width=0.40\textwidth]{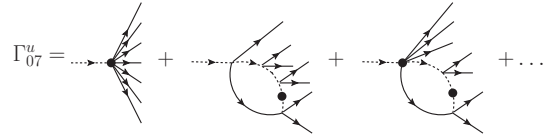}
\caption{Sample of one-loop 8-point functions.  A blob on a vertex represents a vertex that is not present at tree-level. \label{fig:8_point_functions}}
\end{figure}

For $m_{b} = 3$, $\Gamma_{10}^{u}$, $\Gamma_{12}^{u}$ and $\Gamma_{14}^{u}$ are divergent in the $\Lambda \rightarrow \infty$ limit.  Those divergences can be absorbed in $r_{u}$, $r_{v}$, $\lambda$, $\lambda_{pq}$ with $p+q = 3, 5$.  Contrary to the $m_{b} = 2$ case, other divergences appear in higher point functions due to the need to introduce {\em new} tree-level couplings $\lambda_{pq}$ with $p+q = 5$.  For example, the 8-point function $\Gamma_{07}^{u}$ (c.f. Fig.~\ref{fig:8_point_functions}) has the following UV behavior:
\begin{equation}
\Gamma_{07}^{u}(0)  \sim  \lambda_{07} + \lambda^{2}A_{u} \sum_{j=1}^{m_{u}}\left[ \lambda^{2} \frac{C_{07}^{(j)}}{\Lambda^{2}} + \lambda_{14} C_{07}^{(j)} \right]\Lambda^{2j+\epsilon - 6}.
\end{equation}
We see that the $O(\lambda^{4})$ contribution is finite whereas the $O(\lambda^{2}\lambda_{14})$ term diverges.  This divergence can be absorbed by introducing a new coupling $\lambda_{07}$, which in turn produces divergences in higher point functions.  Thus for $m_{b}=3$, an infinite number of divergences appear, necessitating the introduction of a corresponding infinite number of new parameters to make the model finite.

We conclude that the stochastic CARD model is renormalizable at one-loop for $m_{b} \leq 2$ and nonrenormalizable for $m_{b} > 2$.  One-loop renormalizability is a necessary but not sufficient condition for renormalizability at all orders.  A full proof requires the analysis of overlapping divergences, which is beyond the scope of the present paper.  However, since the CARD model has a structure similar to a scalar $\lambda\phi^{4}$ theory, we expect renormalizability to hold to all orders.

Formally, the above result can also be obtained from an analysis of the superficial degree of divergence $D$ of each diagram (e.g. \cite{AlvarezGaume_2012,Schwartz_2014}):
\begin{equation}
D  =  (d_{s}-2)L + 2 - N_{v}y_{v} - N_{u}y_{u}  - 2\sum_{p,q}n_{pq}
\end{equation}
where $L$ is the number of loops, $N_{b}$ the number of noise contractions (c.f.~Fig.~\ref{fig:Feynman_rules}) and $n_{ab}$ the number of vertices with $p$ $U$ legs and $q$ $V$ legs.  For $L=1$, the superficial degree of divergence reduces to $D = 2m_{b} - 2\sum_{pq}n_{pq}$.  By systematically working out the possible combinations of vertices, one can show that $D \geq 0$ for a finite number of one-loop diagrams only when $m_{b} \leq 2$.

Note that the inducing of interactions is selective, since $m_{u}$ and $m_{v}$ are independent free parameters.  Experimentally, the noise for each species can come from different sources (e.g. temperature, luminosity, stirring, etc).  It is thus in principle possible to control the generation of higher order interactions involving only one type of noise.

It is worth pointing out that in general, field theories have definite properties of (non) renormalizability (e.g. Fermi and electroweak theories in particle physics).  The reason for this is that the dimension of spacetime is generally fixed.  In contrast, the power counting parameter responsible for divergences is the effective dimension $d_{\rm eff} = d_{s} - y_{b}$ for stochastic field equations with power law noise.  We argue in the following that this unique feature could be used to probe chemical reactions with externally tunable noise.

\section{Effects of higher order interactions}
\label{sec:Higher_order_interactions}

\begin{figure}
\includegraphics[width=0.40\textwidth]{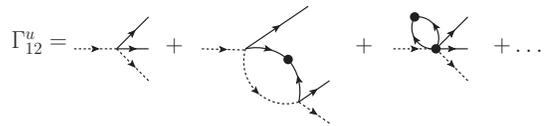}
\caption{Typical corrections to $\lambda$ coming from renormalizable and nonrenormalizable interactions.  \label{fig:Corrections_lambda_higher_order}}
\end{figure}

As we have seen, an infinite number of higher-point functions contribute to the renormalization of the starting model parameters ($D_{b}$, $r_{b}$, $\lambda$) for $m_{b} > 2$.  Fortunately, this does not prevent the model from being predictive.  In the context of effective field theory (e.g. \cite{Burgess_2007}), it is assumed that there is a scale $r_{w} \gg r_{b}$ at which a (macroscopic) model must be replaced by a more fundamental (microscopic) model.  At low wavenumbers (or large scales),  loop corrections coming from these nonrenormalizable interactions are suppressed by the small ratio $(D_{b}|{\bf k}|^{2} + r_{b})/r_{w}$.  For example, typical corrections to the coupling $\lambda$ when $m_{b} = 3$ are given by (c.f. Fig.~\ref{fig:Corrections_lambda_higher_order}):
\begin{eqnarray}
\Gamma_{12}^{u}(k) & \sim & \lambda\left[1 + \left(\frac{\lambda A_{v}(D_{v}r_{u}+D_{u}r_{v}+2D_{v}r_{v})}{D_{v}^{2}(D_{u}+D_{v})^{2}} \right)\ln\Lambda \right. \nonumber \\
                   &      & \left. + \lambda_{14}'\left(\frac{\lambda A_{v}r_{v}}{D_{v}^{3}}\right) \left(\frac{r_{v}}{r_{w}}\right)\ln\Lambda \right] + O\left(\frac{r_{b}^{2}}{r_{w}^{2}}\right), 
\end{eqnarray}
where we write $\lambda_{14} = \lambda_{14}'\lambda^{2}/r_{w}$ (with $\lambda_{14}'$ a dimensionless free parameter), in accordance to the fundamental assumption of effective field theory that there is a hierarchy of scales.  Note that the correction due to the nonrenormalizable interaction is suppressed by $r_{v}/r_{w}$ with respect to the renormalizable one.  It is thus a perfectly controllable expansion at large scales and one can therefore use it for phenomenological purposes.  

As argued in Sect.~\ref{sec:Philosophy_goals}, this hierarchy of nonrenormalizable terms is the low wavenumber manifestation of new chemical pathways evolving at shorter temporal and spatial scales.  Those pathways are macroscopic, in the sense that they only involve $U$'s and $V$'s.  To understand what is happening microscopically (e.g. inside the blob of the first diagram in Fig.~\ref{fig:6_point_functions}), one needs to find a more fundamental model (involving additional species) valid at those small scales.  We explain this idea in the next section.

\section{Effective field theory and chemistry}
\label{sec:EFT_chemistry}

The reaction~(\ref{eq:Gray_Scott_reactions}) involves three chemicals reacting together at a ``point'' (or more precisely a coarse-grained fluid cell) represented by cubic interactions in Eqs.~(\ref{eq:Gray_Scott_equations_1})-(\ref{eq:Gray_Scott_equations_2}).  It is natural to think that this reaction can be written as two reactions involving an intermediate species evolving on shorter scales.  In what follows we argue that this idea is captured by effective field theory, thus allowing the building of microscopic chemical models from macroscopic ones. 

\begin{figure}
\includegraphics[width=0.35\textwidth]{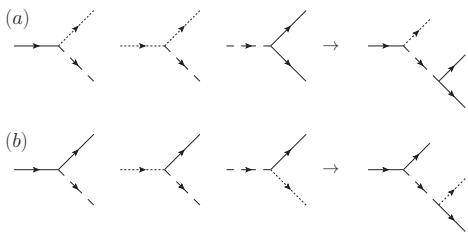}
\caption{Possible sets of cubic vertices leading to the tree-level quartic vertices shown in Fig.~\ref{fig:Feynman_rules}.  Dashed lines correspond to an intermediate species $W$ evolving on shorter scales. \label{fig:Possible_cubic_vertices}}
\end{figure}

To illustrate this, consider the case where $m_{b} = 1$ for simplicity. Let us construct a (microscopic) model with cubic interactions involving a new species $W$ acting on scales $r_{w} \gg r_{b}$.  This model has to match with the macroscopic model at large scales, up to corrections suppressed by powers of the large scale $r_{w}$ \footnote{This is different from the coarse-graining approach in condensed matter physics.  In coarse-graining, one starts from a microscopic model and averages it to obtain its large scale properties.  There is no change in degrees of freedom (e.g. number of particle species, number of chemical species, etc) during this procedure.  In fine-graining, one typically observes such a change in degrees of freedom (e.g. going from (U,V) to (U,V,W), general mechanism for methane oxidation or any other chemical mecanism~\cite{Steinfeld_etal_1998}).}.  There are two sets of cubic interactions that satisfy this criterion (see Fig.~\ref{fig:Possible_cubic_vertices}).  The corresponding evolution equations for set (a) are (set (b) is done in a similar way):
\begin{eqnarray}
\label{eq:Microscopic_model_1}
\partial_{t} V & = & D_{v}\nabla^{2}V - r_{v}V + g_{1}UW + \eta_{v}(x), \\
\label{eq:Microscopic_model_2}
\partial_{t} U & = & D_{u}\nabla^{2}U - r_{u}U + g_{2}UW + \eta_{u}(x) + f, \\
\label{eq:Microscopic_model_3}
\partial_{t} W & = & D_{w}\nabla^{2}W - r_{w}W + g_{3}V^{2} + \eta_{w}(x).
\end{eqnarray}
At large scales, temporal and spatial variations in $W$ can be neglected in Eq.~(\ref{eq:Microscopic_model_3}), leading to $W = g_{3}V^{2}/r_{w} + O(1/r_{w}^{2})$.  Plugging this result in Eqs.~(\ref{eq:Microscopic_model_1})-(\ref{eq:Microscopic_model_2}), we recover the CARD equations~(\ref{eq:Gray_Scott_equations_1})-(\ref{eq:Gray_Scott_equations_2}) with the identifications $\lambda = g_{1}g_{3}/r_{w} = -g_{2}g_{3}/r_{w}$, up to $O(1/r_{w}^{2})$ corrections.  This is an example of matching, as explained in Sect.~\ref{sec:Philosophy_goals}.  The corrections enable the distinction between different microscopic models.  Note that the above procedure is very similar to the one in particle physics leading to the Fermi theory of the weak interactions from the electroweak theory \footnote{More precisely, since the CARD model is renormalizable for $m_{b}=1$, the procedure described above is more akin to finding a high energy model that would give the (renormalizable) Standard Model at low energies, plus small corrections to explain neutrino masses.  We use the simple $m_{b}=1$ case only to illustrate the low energy expansion of effective field theory and its link with chemistry.  The discussion of the $m_{b}=3$ case, including all possible nonrenormalizable terms, is beyond the scope of the present paper and is left for a future publication.}.


From a diagrammatic point of view, both sets of cubic interactions lead to the CARD equations, though only one has an interpretation in terms of chemistry.  The chemical reactions corresponding to set (a)  are:
\begin{equation}
\mbox{U + W}  \stackrel{g_{1}}{\rightarrow}  \mbox{V + W},\;\;\;\;\; \mbox{2V} \stackrel{g_{3}}{\rightarrow} \mbox{W},\;\;\;\;\; \mbox{W} \stackrel{r_{w}}{\rightarrow} 2\mbox{V}\; ,
\end{equation}
up to $O(1/r_{w}$) corrections.  These reactions are equivalent to Eq.~(\ref{eq:Gray_Scott_reactions}).  For set (b), it can be shown by inspection that it is not possible to write down a set of chemical reactions leading to Eq.~(\ref{eq:Gray_Scott_reactions}).  In this simple case, stoichiometry (related to conservation laws and thus symmetries for chemical reactions) and the law of mass action dictate the appropriate microscopic chemical model.

\section{Discussion}
\label{sec:Discussion}

From the above results, the following picture emerges.  Noise  can be viewed as a microscopic probe that can be externally tuned.  Indeed, there is a microscopic scale $\Delta x \sim (A_{u}r_{u}/f^{2})^{1/m_{u}}$ at which noise is as important as diffusion, decay and interactions (obtained by comparing terms in Eqs.~(\ref{eq:Gray_Scott_equations_1})-(\ref{eq:Gray_Scott_equations_2})).  Increasing noise has the effect of bringing fluctuation-dominated physics into the foreground.  This manifests itself in the generation of additional terms to the reaction part  of the reaction-diffusion system.  Since noise fluctuations are represented by loops in Feynman diagrams, increasing the noise exponent  (for a power law noise) has the effect of inducing new interactions (and therefore new chemical pathways) not otherwise present at tree-level.  For example, the renormalizable induced interaction $\Gamma_{03}$ corresponds to the inverse reaction 3V $\rightarrow$ U + 2V (other chemical pathways are possible too) and might help in restoring detailed balance at small scales.  Nonrenormalizable interactions ($\Gamma_{pq}$ with $p+q \geq 5$) correspond to potentially new chemical reactions at small scales, suppressed by powers of the small ratio (probing scale)$^{-1}$/(``new chemistry'' scale)$^{-1}$, where the ``new chemistry'' scale must be larger than the smallest microscopic distance scale (dictated by the chemical constituents themselves).  These new induced interactions can possibly be included in a more fundamental microscopic model involving intermediate species using the tools of effective field theory, as we illustrate in our simple example.  We are thus effectively probing the innards of the equations with noise.

Coarse-graining implies a loss of information due to averaging.  Here, we have explored the inverse problem, namely fine-graining the equations in order to understand the underlying microscopic model.  This process is not unique, and additional information (symmetries, experimental input, etc) is necessary to better pin down the correct microscopic model.  Our simple example (with two possible models represented by two sets of cubic vertices) is an illustration of this fact.

\section{Conclusion}

This paper presents a systematic attempt to fine-grain the stochastic CARD model.  As explained in Sect.~\ref{sec:Philosophy_goals}, this fine-graining can be naturally couched in the language of renormalization and effective field theory.  We show how the interplay between noise and renormalizability leads to the appearance of new nonrenormalizable interactions interpreted as new chemical pathways.

The ideas presented in this paper will be developed further from a theoretical, numerical and experimental point of view in future publications.  We hope that this work may bring noise control and the powerful machinery of (effective) field theory to the analysis and construction of chemical models, and may ultimately help in the understanding of more complex chemical and biological systems.

\begin{acknowledgments}
The authors thank M. Gell-Mann, G. Goshal, R. Hausmann, C. Hidalgo, D. Hochberg, A. Mu\~{n}uzuri, J. Szymanski and  E. Szabo for useful discussions and Repsol S.A. for their support.
\end{acknowledgments}

\bibliography{renormalizibility_GS}

\end{document}